\documentclass[prd ]{revtex4}
\usepackage{caption}
\usepackage{amsmath}
\usepackage{graphicx}
\usepackage{epsfig}
\usepackage{dcolumn}
\usepackage{bm}
\usepackage{slashed}
\usepackage{amsfonts}
\usepackage{amssymb}%
\providecommand{\U}[1]{\protect\rule{.1in}{.1in}}
\begin{document}
\title{ A relativistic formulation of the de la Pe\~{n}a-Cetto \\stochastic quantum mechanics. \\}
\author{Alfredo Gonz\'{a}lez Lezcano$^{a}$ and Alejandro Cabo Montes de Oca$^{b}$}
\affiliation{$^{a}$ Departamento de F\'{\i}sica, Universidad de Pinar del R\'{\i}o,}
\affiliation{Pinar del R\'{\i}o, Cuba,\\}
\affiliation{$^{b}$ Instituto de Matem\'{a}tica, Cibern\'{e}tica y F\'{\i}sica, }
\affiliation{Calle E, No. 309, e/ 13 y 15, Vedado, La Habana, Cuba. }

\begin{abstract}
\noindent A covariant generalization of a non-relativistic stochastic quantum
mechanics introduced by de la Pe\~{n}a and Cetto is formulated. The analysis
is done in space-time and avoids the use of a non-covariant time evolution
parameter in order to search for Lorentz invariance. The covariant form of the
set of iterative equations for the joint coordinate and momentum distribution
function $Q(x,p)$ is derived and expanded in power series of the charge of the
particle. Then, solutions of the zeroth order in the charge of the iterative
equations for $Q(x,p)$ are considered. For them, it follows that the
space-time probability density $\rho(x)$ and the function $S(x)$ which
gradient defines the mean value of the momentum at the space time point $x$,
define a complex function $\psi(x)$ which exactly satisfies the Klein-Gordon
(KG) equation. It is argued that when the KG solution is either of positive or
negative energy, the total number of particles conserves in the random motion.
It follows that the total number of particles conserves when $\psi(x)$
composed alternatively of pure positive or negative energy modes. In addition,
solutions for the joint distribution function in lowest order, satisfying the
positive condition are also found. They suggests a possible link with the
Couder's experiments on droplet movements over oscillating liquid surfaces.

\end{abstract}
\maketitle

\section{Introduction}

The search for stochastic descriptions of quantum mechanics and quantum fields
theory has a large history. By example, in the works
\cite{marshall1,marshall2,marshall3,boyer,penacetto,franca}, it was considered
that the random forces which determine the phase space density of particle,
were given in a stochastic and relativistic invariant way. This defined the so
called Stochastic Electrodynamics (SQED) . Specifically, these forces were
generated by an electromagnetic field configuration, obtained by exciting all
the oscillation modes with one half a quanta of energy. In addition, the
phases of the photon modes were assumed to be stochastically defined within
the interval $(0,2\pi)$ in a uniform way. This force had been argued to be
invariant under Lorentz transformations \cite{marshall1,boyer}. In addition,
in reference \cite{penacetto}, it was argued that the stochastic motion guided
by those forces, when taken in the non-relativistic limit, leads to the
Schrodinger equation. More recently, and with the purpose of start considering
a generalization of the results in Ref.\cite{penacetto}, in reference
(\cite{stoch}), it was suggested that random forces, showing the same
statistical distribution in all Lorentz frames, can be expected to imply the
satisfaction of the Klein-Gordon equation, for which the non-relativistic
limit is the Schrodinger one.

This work is devoted to present a derivation of \ a relativistic version of
the stochastic electrodynamics. That is, we intend to relativistically
generalize the discussion done in Ref. \cite{penacetto}. For this purpose we
start from the relativistic formulation of the kinetic equations given in Re
\cite{deGroot}. The implementation of the stochastic force is assumed to
coincide with the one employed in the stochastic electrodynamics \cite{boyer}.
Then, a formula for the equation satisfied by the fluctuation independent
space-momenta joint distribution function is derived. It directly generalizes
the expression obtained in \cite{penacetto} for the non-relativistic limit. A
formula for the joint distribution is derived.

Then, the solution of the equations for the joint particle distribution is
searched as expanded in a power series in the squared particle charge.
\ Further, it is shown the existence of solutions for the joint distribution
in the zeroth order in the charge. This implies that the density of particles
at a given space time point is defined as a square of complex function
satisfying the Klein-Gordon equation, in this considered interaction free
approximation. This indicates that the Klein-Gordon equation (or its
non-relativistic limit the Schrodinger one) can be described as satisfying
some of the equations of SQED in the first approximations. It can be cited
that recently, in Ref. \cite{grossing} a derivation was presented of the
Klein-Gordon equation, from a modified classical Hamilton-Jacobi equation for
a particle interacting with random background forces. The present discussion
independently support this conclusion through a generalization of the
non-relativistic kinetic discussion of the non-relativistic limit done in
\cite{penacetto}. The presentation also clarifies the role of the positive and
negative energy solutions, by showing that both of them separately imply the
conservation of the total number of particles in the stochastic motions,
assumed that external electromagnetic fields are absent.

In addition an attempt is done for finding solutions for the waves $\psi(x)$
defining a joint distribution function properly only on the mass shell
$p^{2}-m^{2}$ of the classical particles. It follows, that no wave function
within the class investigated can satisfy this condition. However, a special
circumstance could help to overcome this negative outcome: The KG equation,
when seen as theory of particles lacks a standard definition for the position
operator having eigen-functions like the Dirac's Delta function. The position
operator for this theory shows Gaussian like spatial behavior with non
vanishing values within an spatial neighborhood of the size of the Compton
wavelength of the particle \ref{schweber}. Then, it looks reasonable that a
clear interpretation of the $\widetilde{Q}^{0}(x,p)$\ as describing particles
with a well defined position $x,$ can run in troubles. Thus, one can imagine
that a proper modification of the kinetic equations to take account of an
extensive nature of the particles could lead to a consistent equivalence of
the SQED with QM and QFT. Assumed that above mentioned difficulty can be
surmounted, an interesting extension of the work could seem feasible. For
this, after including an external electromagnetic field, it seem possible to
develop a picture in which both types of particles move randomly: one kind of
them guided by the positive energy solutions and the other one (with opposite
charges) moves as driven by the negative energy waves. The development of such
a picture is an interesting envisioned extension of the work.

In the work, we also present solutions for the joint distribution function
$Q(x,p)$ which have positive values in all the phase space points. This is
attained by employing Yukawa like localized solutions of the Klein-Gordon
equation in the construction. These distributions suggests the possibility of
describing sets of extended moving particles, showing the surprising
experimental properties shown by droplets moving over oscillating liquid
surfaces \cite{couder,couder1}. The idea comes from the suspected possibility
of approximately representing "moving" Yukawa like solutions as wavepackets of
massive propagating modes with constant velocity $\ v=\frac{p}{\sqrt
{p^{2}+m^{2}}}.$ The investigation of the scattering properties of such
solutions on two slits screens will be investigated elsewhere.

In Section II we introduce the basic notions of the relativistic kinetic
theory. Next, in Section III, the relativistically invariant equations for the
mean value of the distribution and its random fluctuations are written.
Further, in Section IV, the momentum Fourier transform of the mean joint
distribution is introduced and the equations for it, are written. Section V,
after considering the equations following in the first order zeroth
approximation in the coupling, exposes how the solutions of the Klein-Gordon
equation define corresponding solutions of the relativistic kinetic equations,
determining a possible joint distribution function in the assumed zeroth order
in the charge. Next, Section VI discusses how the defined solutions determine
particle distributions which conserve the total number of particles when the
KG waves are assumed to be alternatively as positive or negative energy modes.
Further, in Section VII , firstly, the mentioned search of free joint
distribution functions being properly defined only on the mass shell of the
particles is exposed. Finally, we present the solutions for the joint
distribution function satisfying the positiveness condition and argue about
its possible links with the Couder's experiments \cite{couder,couder1}.

The results are reviewed and commented in the Summary section.

\section{The equation for the joint distribution function}

Let us start by writing the relativistic invariant equation for the density of
points in phase space $R(x,p)$ for an ensemble of massive particles all
evolving under the action of a stochastic 4-force $F^{\mu}(x,p)$ which was
derived in Ref. \cite{deGroot}%
\begin{equation}
p^{\mu}\frac{\partial}{\partial x^{\mu}}R(x,p)+m\text{ }F^{\mu}(x,p)\frac
{\partial}{\partial p^{\mu}}R(x,p)=0. \label{equation}%
\end{equation}

The 4-coordinates $x^{\mu}$ will be considered in the metric
\begin{equation}
g^{\mu\nu}=\left[
\begin{tabular}
[c]{llll}%
$1$ & $0$ & $0$ & $0$\\
$0$ & $-1$ & $0$ & $0$\\
$0$ & $0$ & $-1$ & $0$\\
$0$ & $0$ & $0$ & $-1$%
\end{tabular}
\ \right]  ,
\end{equation}
where the natural system of coordinates will be also employed, in which the
light velocity $c=1$ and the time is the $x^{0}$ coordinate and the Planck
constant $\hbar=1$. The four momentum as usual, is given in terms of the
3-velocity as
\begin{equation}
p^{\mu}=\frac{m(1,\overrightarrow{v})}{\sqrt{1-\overrightarrow{v}^{2}}}.
\end{equation}

In order to simplify the discussion, we will firstly consider that the
external force vanishes. The particle density $n(\overrightarrow{x},t)$ and
the particle flow $\overrightarrow{j}(\overrightarrow{x},t)$ in this
relativistically invariant case have the form
\begin{align}
N^{\mu}(x)  &  =(n(\overrightarrow{x},t),\overrightarrow{j}(\overrightarrow{x}%
,t))\nonumber\\
&  =\int\frac{d\overrightarrow{p}}{p^{0}}p^{\mu}\text{ }R(x,p),
\end{align}
in which the integration is over all the 3-momenta. In general the conventions
defined in Ref. \cite{deGroot} will be employed. It will be assumed that the
momenta values are defined on the mass-shell
\[
p^{2}-m^{2}=0.
\]

In the present work, the force will assumed to be stochastically defined as in
reference \cite{boyer}. That force had been argued to be invariant under
Lorentz transformations \cite{marshall1,marshall2,marshall3,boyer}. Therefore,
the equation (\ref{equation}) also becomes relativistically invariant in form.

As remarked before in reference \cite{penacetto}, it was argued that the
stochastic motion guided by that force, when taken in the non-relativistic
limit, leads to the Schrodinger equation in the first steps of an iterative
process of solution of the equations for the non relativistic distribution
function. Therefore, as it was argued in reference \cite{stoch}, it can be
suspected that the relativistic invariant motions determined by
(\ref{equation}) could be related with the satisfaction of the Klein-Gordon
equation, for which the non-relativistic limit is the Schrodinger one. This
work is devoted to investigate this possibility. In order to make the
discussion clearer let us argue in the next section that in the
non-relativistic limit, the stochastic equations reproduce the ones employed
in Ref. \cite{penacetto}.

\subsection{The non-relativistic limit of the equation}

In this case since $\frac{\overrightarrow{v}^{2}}{c^{2}}\ll1$ the momentum and
the external force can be approximately given by
\begin{equation}
p^{\mu}=m(1,\overrightarrow{v}).
\end{equation}
Then, after considering $\overrightarrow{v}=\frac{\overrightarrow{p}}{m}$ the
equation (\ref{equation}) reduces to
\begin{align}
\frac{\partial}{\partial x^{0}}R(x,\overrightarrow{p})+\frac
{\overrightarrow{p}}{m}\cdot\frac{\partial}{\partial\overrightarrow{x}%
}R(x,\overrightarrow{p})+\overrightarrow{F}(x,p)\frac{\partial}{\partial
\overrightarrow{p}}R(x,\overrightarrow{p})  &  =\nonumber\\
\frac{\partial}{\partial x^{0}}R(x,\overrightarrow{p})+\frac{\partial
}{\partial\overrightarrow{x}}(\frac{\overrightarrow{p}}{m}\cdot
R(x,\overrightarrow{p}))+\overrightarrow{F}(x,p)\cdot\frac{\partial}%
{\partial\overrightarrow{p}}R(x,\overrightarrow{p})  &  =0.
\end{align}

Further, assuming that $\overrightarrow{F}(x,p)=\overrightarrow{F}(x),$ in
other words that the force is independent of the momentum, leads to
\begin{equation}
\frac{\partial}{\partial x^{0}}R(x,\overrightarrow{p})+\frac{\partial
}{\partial\overrightarrow{x}}(\frac{\overrightarrow{p}}{m}\cdot
R(x,\overrightarrow{p}))+\frac{\partial}{\partial\overrightarrow{p}}%
\cdot(\overrightarrow{F}(x)\text{ }R(x,\overrightarrow{p}))=0.
\end{equation}
which is the same starting formula employed in \cite{penacetto}.  However, in
order to arrive to this expression it was assumed that the force is not
momentum dependent. However, the stochastic electric force term employed in
\cite{penacetto} is momentum independent, an thus it makes the
non-relativistic equation employed in reference \cite{penacetto} and the one
employed here, equivalent in the non-relativistic limit.

\subsection{The adopted SQED\ relativistic random vacuum forces}

Let us give a precise definition of the relativistic stochastic process under
consideration. Note first that we had omitted a time $t$ argument in the
distribution in order to avoid the use of the non-relativistic invariant
definition of the time. Therefore, the stochastic character of the process
will be implemented by defining a large ensemble of particle trajectories in
the phase space $(x,p).$ Each of these trajectories will be defined by a
solution of the equation (\ref{equation}) for a force given by a random
realization of the relativistic invariant Lorentz force employed in SQED
\cite{boyer}%
\begin{align}
F^{\mu}(x,p)  &  =\frac{q}{m}F_{\nu}^{\mu}(x)p^{\nu}\nonumber\\
&  =q\text{ }f^{\mu}(x,p),\\
f^{\mu}(x,p)  &  =F_{\nu}^{\mu}(x)\frac{p^{\nu}}{m},
\end{align}
in which the stochastic space-time dependent field intensity $F_{\gamma\beta
}(x)$ is given by%
\begin{align}
F_{\gamma\beta}(x)  &  =\partial_{\gamma}A_{\beta}(x)-\partial_{\beta
}A_{\gamma}(x),\\
A_{\beta}(x)  &  =(A_{0}(x),\overrightarrow{A}(x)),\\
\overrightarrow{A}(x)  &  =\sum_{\lambda=1}^{2}\int d\overrightarrow{k}%
\frac{1}{w_{k}}\overrightarrow{\epsilon}(\overrightarrow{k},\lambda
){h(\overrightarrow{k},\lambda)\times}\\
&  \sin(\overrightarrow{k}\cdot\overrightarrow{x}-w_{k}\text{ }x^{0}%
+\theta(\overrightarrow{k},\lambda)),\nonumber
\end{align}
where $w_{k}=|\overrightarrow{k}|$ , $\overrightarrow{\epsilon}%
(\overrightarrow{k},\lambda)$ are two unit polarization vectors associated to
the wave vector $\overrightarrow{k}$ and satisfying%
\begin{equation}
\overrightarrow{\epsilon}(\overrightarrow{k},\lambda)\cdot
\overrightarrow{\epsilon}(\overrightarrow{k},\lambda^{\prime})=\delta
_{\lambda\lambda^{\prime}},\text{ \ \ \ }\overrightarrow{k}\cdot
\overrightarrow{\epsilon}(\overrightarrow{k},\lambda)=0,
\end{equation}
and the number $h$ is defined as satisfying
\begin{equation}
\pi^{2}h^{2}=\frac{1}{2}w_{\overrightarrow{k}}.
\end{equation}

Finally, the phases $\theta(\overrightarrow{k},\lambda)$ are defined as
independent random (one for each value of $(\overrightarrow{k},\lambda)$) and
uniformly distributed in the interval $(0,2\pi)$ \cite{boyer}.

In what follows, in place of the force expression, we will prefer to work with
the defined above force per unit of charge
\begin{align}
f^{\mu}(x,p)  &  =\frac{1}{q}F^{\mu}(x,p)\nonumber\\
&  =F_{\nu}^{\mu}(x)\frac{p^{\nu}}{m}.
\end{align}

\subsection{Definitions for the operators and their kernels in joint
coordinate-momentum space}

We will consider in what follows linear kernels in the space of functions
$PS=\{g(x,p)\}$of the particle coordinates $x$ and momenta $p$, which
explicitly written will make the expression to appear as cumbersome. \ Then,
for any of such kernels $K,$ which action on functions of the space \ $g$ is
defined as%
\begin{equation}
g^{\prime}(x,p)=%
{\displaystyle\int}
dx^{\prime}\text{ }dp^{\prime}\text{ }K(x,p;x^{\prime},p^{\prime})\text{
}g(x^{\prime},p^{\prime}),
\end{equation}
its compact operator expression will be defined according to the following
equivalence rules
\begin{align}
g(x,p)  &  \equiv g,\\%
{\displaystyle\int}
dx^{\prime}\text{ }dp^{\prime}\text{ }K(x,p;x^{\prime},p^{\prime})\text{
}g(x^{\prime},p^{\prime})  &  \equiv\widehat{K}\text{ }g.
\end{align}

The special Delta function kernel $\delta^{(8)}(x-x^{\prime},p-p^{\prime
})=\delta^{(4)}(x-x^{\prime})$ $\delta^{(4)}(p-p^{\prime})$ will be simply
defined as the identity \ $\widehat{I},$ \ which will mean for the kernel
associated to the inverse of $K$, the relation
\[
\widehat{K}^{-1}\widehat{K}=\widehat{K}\text{ }\widehat{K}^{-1}=\widehat{I}.
\]

The local operators, like $\ L=p^{\mu}\frac{\partial}{\partial x^{\mu}}$ and
$f=f^{\mu}(x,p)\frac{\partial}{\partial p^{\mu}}$ \ are also considered as
kernels \ in the usual way
\begin{align}
p^{\mu}\frac{\partial}{\partial x^{\mu}}  &  \rightarrow p^{\mu}\frac
{\partial}{\partial x^{\mu}}\delta^{(8)}(x-x^{\prime},p-p^{\prime}%
)\equiv\widehat{L},\\
f^{\mu}(x,p)\frac{\partial}{\partial p^{\mu}}  &  \rightarrow f^{\mu
}(x,p)\frac{\partial}{\partial p^{\mu}}\delta^{(8)}(x-x^{\prime},p-p^{\prime
})\equiv\widehat{f}.
\end{align}

\section{The equations for the joint distribution function}

Now, we will apply the method of smoothing (See \cite{smooth} and
\cite{penacetto}) in order to reduce the equation (\ref{equation}) to a non
random one for the coordinate-momenta joint distribution function over the
defined ensemble of trajectories. The ensemble is generated by samples of the
stochastic force, generated by the random phases of the electromagnetic modes
$\theta(\overrightarrow{k},\lambda)$ taken for all the values of momenta and
polarization $(\overrightarrow{k},\lambda)$ . For the further analysis, the
distribution function will be decomposed in its average coordinate-momenta
joint distribution $Q(x,\overrightarrow{p})$ and its random fluctuations
$\delta Q(x,\overrightarrow{p})$ as
\begin{equation}
R(x,p)=Q(x,p)+\delta Q(x,p),
\end{equation}
with
\begin{align}
Q(x,\overrightarrow{p})  &  =\widehat{P}\text{ }R(x,p),\\
\delta Q(x,\overrightarrow{p})  &  =(1-\widehat{P})R(x,p),
\end{align}
where the $\widehat{P}$ is a projection operator satisfying $\widehat{P}%
^{2}=1.$ After substituting these expression in equation (\ref{equation}) and
applying alternatively $\widehat{P}$ or $(1-\widehat{P}),$ the following two
equations follow
\begin{align}
L\text{ }Q(x,p)+q\text{ }\widehat{P}\text{ }f^{\mu}(x,p)\frac{\partial
}{\partial p^{\mu}}\delta Q(x,p)  &  =0,\nonumber\\
L\text{ }\delta Q(x,p)+q\text{ }(1-\widehat{P}\text{ })f^{\mu}(x,p)\frac
{\partial}{\partial p^{\mu}}Q(x,p)  &  =0,\label{pair}\\
L  &  =p^{\mu}\frac{\partial}{\partial x^{\mu}}.
\end{align}

Consider now the retarded Green function in the space $PS$ of the differential
operator $L=p^{\mu}\frac{\partial}{\partial x^{\mu}},$ satisfying%
\begin{equation}
p^{\mu}\frac{\partial}{\partial x^{\mu}}G(x,p;x^{\prime},p^{\prime}%
)=\delta^{(4)}(x-x^{\prime})\delta^{(4)}(p-p^{\prime}).
\end{equation}

In terms of its Fourier transform in the two arguments, the Green function $G$
takes the form%
\begin{align}
G(x,p;x^{\prime},p^{\prime})  &  =\int\int\frac{dq}{(2\pi)^{4}}\text{ }%
\frac{dz}{(2\pi)^{4}}\frac{1}{-i\text{ }p^{\mu}q_{\mu}}\exp(-i(x-x^{\prime
})^{\mu}q_{\mu}-i(p-p^{\prime})^{\mu}z_{\mu})\nonumber\\
&  =\int\int\frac{dq}{(2\pi)^{4}}\text{ }\frac{1}{-i\text{ }p^{\mu}q_{\mu}%
}\exp(-i(x-x^{\prime})^{\mu}q_{\mu})\text{ }\delta(p-p^{\prime})\nonumber\\
&  =G(x-x^{\prime}|p^{\prime})\text{ }\delta(p-p^{\prime})\equiv
\widehat{L}^{-1}.
\end{align}

\ It is important to note here, that this expression for $G$ indicates that
the \ derivatives $\frac{\partial}{\partial p^{\mu}}$ do not commute with the
operator $G$, since%
\begin{align}
\frac{\partial}{\partial p^{\mu}}G(x,p;x^{\prime},p^{\prime})  &
=G(x-x^{\prime}|p^{\prime})\text{ }\frac{\partial}{\partial p^{\mu}}%
\delta(p-p^{\prime})\nonumber\\
&  =-G(x-x^{\prime}|p^{\prime})\text{ }\frac{\partial}{\partial p^{\prime\mu}%
}\delta(p-p^{\prime})\nonumber\\
&  =G(x-x^{\prime}|p^{\prime})\text{ }\delta(p-p^{\prime})\frac{\partial
}{\partial p^{\prime\mu}}\nonumber\\
&  +\frac{\partial}{\partial p^{\prime\mu}}(G(x-x^{\prime}|p^{\prime}))\text{
}\delta(p-p^{\prime})\nonumber\\
&  \neq G(x-x^{\prime}|p^{\prime})\text{ }\delta(p-p^{\prime})\frac{\partial
}{\partial p^{\prime\mu}}. \label{noncom}%
\end{align}

Therefore, this non commutativity of the momentum derivative with the
propagator $G$ made difficulty to show in this relativistic case an important
property derived in reference \cite{penacetto}: the quadratic dependence in
the Fourier transform variable $z$ of the momentum $\ p$, of some relevant
quantities in the discussion. This lack of commutativity led us, further ahead
in this work, to consider the expansion in the coupling in the equations, in
place of the expansion in powers of $z$.

Now, acting with the product of $G$ and $p^{\mu}\frac{\partial}{\partial
x^{\mu}}$ on an arbitrary function $g(x,p)$ it follows
\begin{align}
\int\int dx^{\prime}dp^{\prime}G(x,p;x^{\prime},p^{\prime})p^{\prime\mu}%
\frac{\partial}{\partial x^{\prime\mu}}g(x^{\prime},p^{\prime})  &  =\int
dx^{\prime}dp^{\prime}\int\frac{dq}{(2\pi)^{4}}\text{ }\frac{\delta
(p-p^{\prime})}{-i\text{ }p^{\mu}q_{\mu}}\exp(-i(x-x^{\prime})^{\mu}q_{\mu
})\times\nonumber\\
&  p^{\prime\mu}\frac{\partial}{\partial x^{\prime\mu}}g(x^{\prime},p^{\prime
})\nonumber\\
&  =g(x,p),
\end{align}
which implies
\begin{equation}
\int\int dx^{\prime}dp^{\prime}\text{ }p^{\mu}\frac{\partial}{\partial x^{\mu
}}G(x,p;x^{\prime},p^{\prime})\equiv\delta(x-x^{\prime})\delta(p-p^{\prime}).
\end{equation}

As in reference we will now define a compact notation in order to eliminate
the cumbersome appearance determined by the kernel structure of the Green
function. This notation is described by expressing the above relation in the
form%
\[
\int\int dx^{\prime}dp^{\prime}p^{\prime\mu}\frac{\partial}{\partial
x^{\prime\mu}}G(x,p;x^{\prime},p^{\prime})p^{\prime\mu}\frac{\partial
}{\partial x^{\prime\mu}}\equiv\widehat{L}\text{ }\widehat{L}^{-1}%
=\widehat{I}\equiv\delta(x-x^{\prime})\delta(p-p^{\prime}).
\]

The use of these relations written above, after acting with the kernel $G$ at
the left of the relations (\ref{pair}) gives for the average of the
distribution and its random part, the expressions
\begin{align}
Q  &  =-q\text{ }\widehat{L}^{-1}\text{ }\widehat{P}\text{ }\widehat{f}\delta
Q,\\
\delta Q  &  =-q\text{ }\widehat{L}^{-1}[\widehat{I}+q\text{ }\widehat{L}%
^{-1}(I-\widehat{P}\text{ )}\widehat{f}]^{-1}\widehat{P}\text{ }\widehat{f}Q.
\end{align}

Then, the substitution of these relations in the first of the equations
(\ref{pair}) leads to the following equations for the joint distribution
function $Q(x,p)$
\begin{align}
\widehat{L}\text{ }Q  &  =q^{2}\widehat{P}\text{ }\widehat{f}\text{
}\widehat{L}^{-1}[\widehat{I}+q\text{ }\widehat{L}^{-1}(\widehat{I}%
-\widehat{P}\text{ )}\widehat{f}]^{-1}\widehat{f}\text{ }Q,\nonumber\\
&  =q^{2}\widehat{P}\text{ }\widehat{f}\text{ }\widehat{L}^{-1}\sum
_{n=0}^{\infty}(-1)^{n}[q\text{ }\widehat{L}^{-1}(\widehat{I}-\widehat{P}%
\text{ })\widehat{f}]^{n}\text{ }\widehat{f}\text{ }Q,\nonumber\\
&  =q^{2}\widehat{P}\text{ }\widehat{f}\text{ }\widehat{L}^{-1}\sum
_{m=0}^{\infty}(q^{2})^{^{m}}\text{ }[\widehat{L}^{-1}(\widehat{I}%
-\widehat{P}\text{ })\widehat{f}]^{2m}\widehat{f}\text{ }Q.
\end{align}

Therefore, the equation for the joint distribution function can be written in
a compact form, which after expanded in powers of $q^{2},$ takes the form%
\begin{align}
\widehat{L}\text{ }Q  &  =q^{2}\text{ }\widehat{J\text{ }}(q^{2})\text{
}Q,\label{joint}\\
\widehat{J\text{ }}(q^{2})  &  =\sum_{m=0}^{\infty}(q^{2})^{^{m}}\text{
}\widehat{P}\text{ }\widehat{f}\text{ }\widehat{L}^{-1}[\widehat{L}%
^{-1}(\widehat{I}-\widehat{P})\text{ }\widehat{f}]^{2m}\widehat{f}\nonumber\\
&  =\sum_{m=0}^{\infty}(q^{2})^{^{m}}\text{ }\widehat{J\text{ }}^{m}(q^{2}).
\end{align}

In this relation it has been employed that the mean value of an odd number of
the random force functions \cite{boyer,penacetto}\ vanishes. We recall that in
that relation $\widehat{f}$ is the operator corresponding to the kernel
\begin{equation}
\widehat{f}\equiv f^{\mu}(x,p)\frac{\partial}{\partial p^{\mu}}.
\end{equation}

\section{ The momentum Fourier transformed joint distribution function}

Let us perform now the Fourier transformation of the joint distribution over
the momentum variable as follows%
\begin{align}
Q(x,p)  &  =\int dz\text{ }\widetilde{Q}(x,z)\exp(-i\text{ }p^{\mu}z_{\mu}),\\
\widetilde{Q}(x,z)  &  =\int\frac{dp}{(2\pi)^{4}}Q(x,p)\exp(i\text{ }p^{\mu
}z_{\mu}). \label{qz}%
\end{align}
Then, after Fourier transforming the equation (\ref{joint}) the following
equation for $Q(x,p)$ can be written%
\begin{equation}
p^{\mu}\frac{\partial}{\partial x^{\mu}}Q(x,p)=q^{2}\sum_{m=0}^{\infty}%
(q^{2})^{^{m}}\int dx^{\prime}dp^{\prime}\text{ }J^{m}(q^{2})(x,p;x^{\prime
},p^{\prime})\text{ }Q(x^{\prime},p^{\prime}),
\end{equation}
\begin{equation}
\frac{\partial}{i\partial x^{\mu}\partial z_{\mu}}\widetilde{Q}(x,z)=q^{2}%
\sum_{m=0}^{\infty}(q^{2})^{^{m}}\int dx^{\prime}d\text{ }z^{\prime}J\text{
}^{m}(q^{2})(x,\frac{\partial}{i\text{ }\partial z};x^{\prime},\frac{\partial
}{i\text{ }\partial z^{\prime}})\text{ }\widetilde{Q}(x^{\prime},z^{\prime}),
\label{general}%
\end{equation}
with \ the operator $J$ $^{m}(q^{2})(x,\frac{\partial}{i\text{ }\partial
z};x^{\prime},\frac{\partial}{i\text{ }\partial z^{\prime}})$ operating in the
space of function of the variables $(x,z)$ is defined by%
\begin{equation}
J^{m}\text{ }(q^{2})(x,\frac{\partial}{i\text{ }\partial z};x^{\prime}%
,\frac{\partial}{i\text{ }\partial z^{\prime}})=\text{ }\left.  \left(
\widehat{P}\text{ }\widehat{f}\text{ }\widehat{L}^{-1}[\widehat{L}%
^{-1}(\widehat{I}-\widehat{P})\text{ }\widehat{f}]^{2m}\widehat{f}\right)
(x,p;x^{\prime},p^{\prime})\right\vert _{p\rightarrow\frac{\partial}{i\text{
}\partial z},\text{ }p^{\prime}\rightarrow\frac{\partial}{i\text{ }\partial
z^{\prime}}}.
\end{equation}

Now, the mean value of a function of the coordinates and momenta at a specific
space-time position $x$ can be written in two forms as
\begin{equation}
\langle A(x,p)\rangle_{x}=\frac{1}{\rho_{t}(x)}%
{\displaystyle\int}
dp\text{ \ }A(x,p)\text{ }Q(x,p)=\frac{1}{\rho_{t}(x)}[A(x,\frac{\partial
}{i\partial z})\text{ }\widetilde{Q}(x,z)]_{z=0},
\end{equation}
where the distribution function in the 3D-space points $\overrightarrow{x}$
and a given time $x_{0}$ is given by
\begin{align}
\rho_{t}(x)  &  =%
{\displaystyle\int}
dp\text{ }Q(x,p)=\sum_{m=0}(q^{2})^{^{m}}\rho^{m}(x)\nonumber\\
&  =\sum_{m=0}(q^{2})^{^{m}}\int dp\text{ }Q^{m}(x,p)\nonumber\\
&  =\sum_{m=0}(q^{2})^{^{m}}\text{ }\widetilde{Q}^{m}(x,0),
\end{align}
in which the general expressions for the distribution function have been
expanded in series of the squared charge as follow%
\begin{align}
Q(x,p)  &  =\sum_{m=0}(q^{2})^{^{m}}Q^{m}(x,p),\\
\text{ }\widetilde{Q}^{m}(x,z)  &  =\sum_{m=0}(q^{2})^{^{m}}\text{
}\widetilde{Q}^{m}(x,z).
\end{align}

Therefore, the general equation (\ref{general}) can be written in the form
\begin{align}
\frac{\partial}{i\text{ }\partial z_{\mu}\partial x^{\mu}}\widetilde{Q}%
^{0}(x,z)  &  =0,\\
\frac{\partial}{i\text{ }\partial z_{\mu}\partial x^{\mu}}\widetilde{Q}%
^{m}(x,z)  &  =q^{2}\sum_{n=0}^{\infty}(q^{2})\int dx^{\prime}d\text{
}z^{\prime}J\text{ }^{n}(q^{2})(x,\frac{\partial}{i\text{ }\partial
z};x^{\prime},\frac{\partial}{i\text{ }\partial z^{\prime}})\text{
}\widetilde{Q}^{m-n-1}(x^{\prime},z^{\prime}),\text{ }m\geq1.
\end{align}

\subsection{\bigskip The interaction free limit $q^{2}\rightarrow0$}

Let us consider now the satisfaction of the first of the iterative equations
in which the random movement had been decomposed. In this case all the
$Q^{m}(x,p)$ for $m\geq1$ will vanish and thus $Q(x,p)=Q^{0}(x,p).$

Then, the total density reduces to%
\begin{align}
\rho_{t}(x)  &  =%
{\displaystyle\int}
dp\text{ }Q^{0}(x,p)=\rho^{0}(x)\nonumber\\
&  =\int dp\text{ }Q^{0}(x,p)\nonumber\\
&  =\widetilde{Q}^{0}(x,0).
\end{align}
The mean values in this limit have the expression%
\begin{equation}
\langle A(x,p)\rangle_{x}=\frac{1}{\rho^{0}(x)}%
{\displaystyle\int}
dp\ A(x,p)Q^{0}(x,p)=\frac{1}{\rho^{0}(x)}[A(x,\frac{\partial}{i\partial
z})\widetilde{Q}^{0}(x,z)]_{z=0},
\end{equation}
in which the distribution function has been expanded in series of the squared charge.

Let us define for what follow
\begin{equation}
\rho(x)=\rho^{0}(x).
\end{equation}

Then, the equation for $\widetilde{Q}^{0}(x,z)$%
\begin{equation}
\frac{\partial}{i\text{ }\partial z_{\mu}\partial x^{\mu}}\widetilde{Q}%
^{0}(x,z)=0,
\end{equation}
can be derived after expanding the exponential in powers of $z.$ \ It is
helpful to write the Fourier transforms in the zero order in $q^{2}$%
\begin{align}
Q^{0}(x,p)  &  =\int\int dz\text{ }\widetilde{Q}^{0}(x,z)\exp(i\text{ }p^{\mu
}z_{\mu}),\\
\widetilde{Q}^{0}(x,z)  &  =\int\int\frac{dp}{(2\pi)^{4}}Q^{0}(x,p)\exp
(-i\text{ }p^{\mu}z_{\mu}).
\end{align}
The mean value formula reduces to
\begin{equation}
\langle A(x,p)\rangle_{x}=\frac{1}{\rho(x)}%
{\displaystyle\int}
dp\text{ \ }A(x,p)\text{ }Q^{0}(x,p)=\frac{1}{\rho(x)}[A(x,\frac{\partial
}{i\partial z})\text{ }\widetilde{Q}^{0}(x,z)]_{z=0},
\end{equation}
which allows to write the lowest order equation in the form
\begin{align}
\frac{\partial}{i\text{ }\partial z_{\mu}\partial x^{\mu}}\widetilde{Q}%
^{0}(x,z)  &  =-\frac{\partial}{\partial x^{\mu}}\int\int\frac{dp}{(2\pi)^{4}%
}Q^{0}(x,p)p^{\mu}\exp(-i\text{ }p\text{ }z)\nonumber\\
&  =\frac{\partial}{\partial x^{\mu}}(\rho(x)\langle p^{\mu}\exp(-i\text{
}p\text{ }z)\rangle_{x}\nonumber\\
&  =0.
\end{align}

We will now introduce new variables $z_{+}$ and $z_{-}$ in substitution of the
variables $x$ and $z$. The change is defined as
\begin{align}
z_{\mu}^{+}  &  =x_{\mu}+\beta\text{ }z_{\mu},\text{ \ \ }z_{\mu}^{-}=x_{\mu
}-\beta\text{ }z_{\mu},\nonumber\\
x_{\mu}  &  =\frac{1}{2}(z_{\mu}^{+}+\beta\text{ }z_{\mu}^{+}),\text{
\ \ }z_{\mu}=\frac{1}{2\beta}(z_{\mu}^{+}-\beta\text{ }z_{\mu}^{+}),
\label{variables}%
\end{align}
and for the derivatives
\begin{align}
\frac{\partial}{\partial z_{\mu}^{+}}  &  \equiv\partial_{+}^{\mu}=\frac{1}%
{2}(\frac{\partial}{\partial x_{\mu}}+\frac{\partial}{\beta\partial\text{
}z_{\mu}}),\text{ \ }\frac{\partial}{\partial z_{\mu}^{-}}\equiv\partial
_{-}^{\mu}=\frac{1}{2}(\frac{\partial}{\partial x_{\mu}}-\frac{\partial}%
{\beta\partial\text{ }z_{\mu}}),\\
\frac{\partial}{\partial x_{\mu}}  &  =(\frac{\partial}{\partial z_{\mu}^{+}%
}+\frac{\partial}{z_{\mu}^{-}}),\text{ \ }\frac{\partial}{\partial z_{\mu}%
}=\beta(\frac{\partial}{\partial z_{\mu}^{+}}-\frac{\partial}{z_{\mu}^{-}}).
\end{align}

These relations allow to derive the identities%
\begin{align}
\frac{\partial^{2}}{\partial z_{\mu}\partial z_{\nu}}  &  =\beta^{2}%
\frac{\partial^{2}}{\partial x_{\mu}\partial x_{\nu}}-2\beta^{2}(\partial
_{-}^{\mu}\partial_{+}^{\nu}+\partial_{+}^{\mu}\partial_{-}^{\nu
})\label{masmenos}\\
\frac{\partial^{2}}{\partial z_{\mu}\partial x^{\mu}}  &  =\beta
(\frac{\partial^{2}}{\partial z_{\mu}^{+}\partial z^{\mu+}}-\frac{\partial
^{2}}{\partial z^{\mu-}\partial z_{\mu}^{-}}).
\end{align}

This last equation permits to write the zeroth order equation for
$\widetilde{Q}^{0}$ in the form\
\begin{equation}
\frac{\partial}{i\text{ }\partial z_{\mu}\partial x^{\mu}}\widetilde{Q}%
^{0}(x,z)=\frac{\beta}{i}(\frac{\partial^{2}}{\partial z_{\mu}^{+}\partial
z^{\mu+}}-\frac{\partial^{2}}{\partial z^{\mu-}\partial z_{\mu}^{-}%
})\widetilde{Q}^{0}(x,z)=0.
\end{equation}

\subsection{Satisfying the higher order equations}

In order to\ directly satisfy the set of equations for $n\geq3$ let us assume
$\widetilde{Q}^{0}(x,z)$ in the form%
\begin{equation}
\widetilde{Q}^{0}(x,z)\equiv\widetilde{Q}^{0}(z^{+},z^{-})=\Psi^{\ast}%
(z^{+})\Psi(z^{-}).
\end{equation}

Then, substituting this form in the zeroth equation leads to%
\begin{equation}
\frac{1}{\Psi^{\ast}(z^{+})}\frac{\partial^{2}}{\partial z_{\mu}^{+}\partial
z^{\mu+}}\Psi^{\ast}(z^{+})-\frac{1}{\Psi(z^{-})}\frac{\partial^{2}}{\partial
z_{\mu}^{-}\partial z^{\mu-}}\Psi(z^{-})=0.
\end{equation}

But this relation is directly satisfied if $\Psi$ obey the linear equation for
any argument $u$ and fixed value of the parameter $M$
\begin{equation}
\frac{\partial^{2}}{\partial u^{\mu}\partial u_{\mu}}\Psi(u)-M^{2}\Psi(u)=0.
\end{equation}

But, the satisfaction of this equation implies
\begin{equation}
\frac{\partial}{i\partial x^{\mu}\text{ }\partial z_{\mu}\partial z_{\mu_{1}%
}\partial z_{\mu_{2}}...\partial z_{\mu_{n}}}\widetilde{Q}^{0}(x,z)=0,\text{
}m=1,2,...\infty, \label{higher}%
\end{equation}
a condition which will be helpful in the further discussion. \

Thus, we had been able to find a solution of the equation describing the free
approximation of the random process. The equation for $\widetilde{Q}^{0}$ can
be also written as
\begin{align}
\frac{\partial}{i\partial x^{\mu}\text{ }\partial z_{\mu}}\widetilde{Q}%
^{0}(x,z)  &  =-\frac{\partial}{\partial x^{\mu}\text{ }}\int\int\frac
{dp}{(2\pi)^{4}}Q^{0}(x,p)\text{ }p^{\mu}\exp(-i\text{ }p^{\mu}z_{\mu
})\nonumber\\
&  =-\frac{\partial}{\partial x^{\mu}\text{ }}(\rho(x)\langle\text{ }p^{\mu
}\exp(-i\text{ }p^{\mu}z_{\mu})\rangle_{x})=0,
\end{align}
which after expanding the exponential in powers of $\ p.z$ \ gives the
following set of equations
\begin{align}
\frac{\partial}{\partial x^{\mu}}(\rho(x)\langle p^{\mu}\rangle_{x})  &  =0,\\
\frac{\partial}{\partial x^{\mu}}(\rho(x)\langle p^{\mu}p^{\nu}\rangle_{x})
&  =0,\\
\frac{\partial}{\partial x^{\mu}}(\rho(x)\langle p^{\mu}p^{\mu_{1}}p^{\mu_{2}%
}...p^{\mu_{n}}\rangle_{x})  &  =0,\text{ }n=1,2,...\infty.
\end{align}

The last of these relations is directly implied by equation (\ref{higher}).
Thus, let us study in what follows the satisfaction of the first two equations
after the adopted ansatz
\[
\widetilde{Q}^{0}(x,z)=\Psi^{\ast}(z_{\mu}^{+})\Psi(z_{\mu}^{-}).
\]

We will use now
\begin{equation}
\langle p^{\mu}p^{\nu}\rangle_{x}=-\frac{1}{\widetilde{Q}^{0}(x,0)}%
[\frac{\partial}{\partial z_{\mu}\partial z_{\nu}}\text{ }\widetilde{Q}%
^{0}(x,z)],
\end{equation}
and the general relations
\begin{equation}
\frac{\partial}{\partial z_{\mu}}(\frac{1}{\widetilde{Q}^{0}(x,z)}%
\frac{\partial}{\partial z_{\mu}}\text{ }\widetilde{Q}^{0}(x,z))=-\frac
{1}{(\widetilde{Q}^{0}(x,z))^{2}}\frac{\partial}{\partial z_{\nu}%
}\widetilde{Q}^{0}(x,z)\text{ }\frac{\partial}{\partial z_{\mu}}%
\widetilde{Q}^{0}(x,z)+\frac{1}{\widetilde{Q}^{0}(x,z)}\frac{\partial
}{\partial z_{\nu}\partial z_{\mu}}\widetilde{Q}^{0}(x,z),
\end{equation}
which after evaluated in $z=0$ permits to write
\begin{align}
\langle p^{\mu}p^{\nu}\rangle_{x}  &  =\langle p^{\mu}\rangle_{x}\langle
p^{\nu}\rangle_{x}-[\frac{\partial^{2}}{\partial z_{\mu}\partial z_{\nu}}%
\ln\text{ }\widetilde{Q}^{0}(x,z))]_{z=0}\nonumber\\
&  =\langle p^{\mu}\rangle_{x}\langle p^{\nu}\rangle_{x}-\beta^{2}%
[\frac{\partial^{2}}{\partial x_{\mu}\partial x_{\nu}}\ln\text{ }%
\widetilde{Q}^{0}(x,z))]_{z=0}+\sigma^{\mu\nu},\\
\sigma^{\mu\nu}  &  =2\beta^{2}\left[  (\partial_{-}^{\mu}\partial_{+}^{\nu
}+\partial_{+}^{\mu}\partial_{-}^{\nu})\ln\widetilde{Q}^{0}\text{ }\right]
_{z=0},
\end{align}
where it was used relation (\ref{masmenos}).

But, the assumed form of the zeroth order distribution allows also to find
\bigskip%
\begin{align}
\sigma^{\mu\nu}  &  =2\beta^{2}\left[  (\partial_{-}^{\mu}\partial_{+}^{\nu
}+\partial_{+}^{\mu}\partial_{-}^{\nu})\ln\widetilde{Q}^{0}\text{ }\right]
_{z=0}\nonumber\\
&  =2\beta^{2}\left[  (\partial_{-}^{\mu}\partial_{+}^{\nu}+\partial_{+}^{\mu
}\partial_{-}^{\nu})(\ln\Psi^{\ast}(z^{+})+\ln\Psi(z^{-})\text{ }\right]
_{z=0}\nonumber\\
&  =0.
\end{align}

Thus, the resting two equations which remaining to be verified in their
compatibility with the ansatz can be written. as%
\begin{align}
\frac{\partial}{\partial x^{\mu}}(\rho(x)\langle p^{\mu}\rangle_{x})  &
=0,\nonumber\\
\frac{\partial}{\partial x^{\mu}}{\large [}\rho(x)(\langle p^{\mu}\rangle
_{x}\langle p^{\nu}\rangle_{x}-\beta^{2}[\frac{\partial^{2}}{\partial x_{\mu
}\partial x_{\nu}}\ln\text{ }\widetilde{Q}^{0}(x,z))]_{z=0}{\large ]}  &  =0.
\label{two}%
\end{align}
\bigskip

\section{The Klein-Gordon equation in the non-interacting limit}

The $q^{2}=0$, will be called the "non interacting or free approximation". In
this section we will study the compatibility of the two equations (\ref{two})
with the expression assumed for the zeroth order joint distribution function.
After checking this, it will follow that in this free limit, the equations
admit solutions for joint distribution function which are defined by waves
solving the Klein-Gordon equations. This result directly generalizes the
derivation of the Schrodinger equation in reference \cite{penacetto}.

Firstly, let us search for solutions of the set of the two equations
\begin{align}
\frac{\partial}{\partial x^{\mu}}(\rho(x)\langle p^{\mu}\rangle_{x})  &
=0,\label{ec1}\\
\frac{\partial}{\partial x^{\mu}}{\large [}\rho(x)\langle p^{\mu}\rangle
_{x}\langle p^{\nu}\rangle_{x}-\beta^{2}\rho(x)[\frac{\partial^{2}}{\partial
x_{\mu}\partial x_{\nu}}\ln\rho(x)\text{ }{\large ]}  &  =0 , \label{ec2}%
\end{align}
where it has been substituted $\rho(x)=\widetilde{Q}(x,0).$ Further, let
search for solutions in which the mean momentum value at a given space-time
point $x$ is defined by%
\begin{equation}
\langle p^{\mu}\rangle_{x}=\frac{\partial}{\partial x_{\mu}}S(x).
\end{equation}

After substituting in (\ref{ec1}) and (\ref{ec2}), it follows%
\begin{align}
\frac{\partial}{\partial x^{\mu}}[\rho(x)\frac{\partial}{\partial x_{\mu}%
}S(x)]  &  =0,\\
\frac{\partial}{\partial x^{\mu}}{\large [}\rho(x)\frac{\partial}{\partial
x_{\mu}}S(x)\frac{\partial}{\partial x_{\nu}}S(x)-\beta^{2}\rho(x)[\frac
{\partial^{2}}{\partial x_{\mu}\partial x_{\nu}}\ln\rho(x)\text{ }{\large ]}
&  =0.
\end{align}

The second equation can be transformed as follows
\begin{align}
\frac{\partial}{\partial x_{\mu}}S(x)\frac{\partial}{\partial x^{\mu}\partial
x_{\nu}}S(x)-\frac{\beta^{2}}{\rho(x)}\frac{\partial}{\partial x^{\mu}}%
\rho(x)[\frac{\partial^{2}}{\partial x_{\mu}\partial x_{\nu}}\ln\rho(x)\text{
}{\large ]}  &  =0,\\
\frac{\partial}{\partial x_{\mu}}{\Large (}\frac{1}{2}\frac{\partial}{\partial
x_{\mu}}S(x)\frac{\partial}{\partial x^{\mu}}S(x){\Large )}-\frac{\beta^{2}%
}{\rho(x)}\frac{\partial}{\partial x^{\mu}}{\Large (}\rho(x)\frac{\partial
^{2}}{\partial x_{\mu}\partial x_{\nu}}\ln\rho(x){\Large )}  &  =0.
\end{align}

But, the density dependent term can be expressed as the divergence of a vector
(as the first term also is) as follows
\begin{align}
\frac{1}{\rho(x)}\frac{\partial}{\partial x^{\mu}}{\Large (}\rho
(x)\frac{\partial^{2}}{\partial x_{\mu}\partial x_{\nu}}\ln\rho(x){\Large )}
&  =\frac{\partial}{\partial x^{\nu}}{\Large (}\frac{1}{2}\frac{\partial
}{\partial x^{\mu}}\ln\rho(x)\frac{\partial}{\partial x_{\mu}}\ln
\rho(x){\Large +}\rho(x)\frac{\partial^{2}}{\partial x_{\mu}\partial x^{\mu}%
}\ln\rho(x){\Large )}\nonumber\\
&  =\frac{\partial}{\partial x^{\nu}}{\Large (-}\frac{1}{2}\frac{\partial
}{\partial x^{\mu}}\ln\rho(x)\frac{\partial}{\partial x_{\mu}}\ln
\rho(x){\Large +}\frac{1}{\rho(x)}\frac{\partial^{2}}{\partial x_{\mu}\partial
x^{\mu}}\rho(x){\Large )}.
\end{align}

Therefore, the following relation arises
\begin{equation}
\frac{\partial}{\partial x_{\nu}}{\Large (}\frac{1}{2}\frac{\partial}{\partial
x_{\mu}}S(x)\frac{\partial}{\partial x^{\mu}}S(x)+\frac{\beta^{2}}{2}%
\frac{\partial}{\partial x^{\mu}}\ln\rho(x)\frac{\partial}{\partial x_{\mu}%
}\ln\rho(x)-\beta^{2}{\Large (}\frac{1}{\rho(x)}\frac{\partial^{2}}{\partial
x_{\mu}\partial x^{\mu}}\rho(x){\Large ))}=0.
\end{equation}

This equation implies,
\begin{equation}
\frac{1}{2}\frac{\partial}{\partial x_{\mu}}S(x)\frac{\partial}{\partial
x^{\mu}}S(x)+\frac{\beta^{2}}{2}\frac{\partial}{\partial x^{\mu}}\ln
\rho(x)\frac{\partial}{\partial x_{\mu}}\ln\rho(x)-\beta^{2}{\Large (}\frac
{1}{\rho(x)}\frac{\partial^{2}}{\partial x_{\mu}\partial x^{\mu}}%
\rho(x){\Large )}=ctc.
\end{equation}

Now, if we fix the constant to be a positive value, given by $ctc=m^{2}$,
\ the searched joint distribution function satisfies the two equations
\begin{align}
\frac{\partial}{\partial x^{\mu}}[\rho(x)\frac{\partial}{\partial x_{\mu}%
}S(x)]  &  =0,\label{a}\\
\frac{1}{2}\frac{\partial}{\partial x_{\mu}}S(x)\frac{\partial}{\partial
x^{\mu}}S(x)+\frac{\beta^{2}}{2}\frac{\partial}{\partial x^{\mu}}\ln
\rho(x)\frac{\partial}{\partial x_{\mu}}\ln\rho(x)-\beta^{2}{\Large (}\frac
{1}{\rho(x)}\frac{\partial^{2}}{\partial x_{\mu}\partial x^{\mu}}%
\rho(x){\Large )}  &  =m^{2}. \label{b}%
\end{align}

\subsection{The satisfaction of the Klein-Gordon equation}

Consider now expressing the KG equation for a complex scalar field $\phi(x)$
in terms of the phase function $S(x)$ and a positive density $\rho(x)$ defined
as follows
\begin{equation}
\phi(x)=(\rho(x))^{\frac{1}{2}}\exp(i\text{ }S(x)).
\end{equation}

Therefore, substituting in the KG equation it follow%
\begin{align*}
(\partial^{2}+m^{2})\phi(x)  &  =0,\\
(\frac{\partial^{2}}{\partial x_{\mu}\partial x^{\mu}}+m^{2})(\rho
(x))^{\frac{1}{2}}\exp(i\text{ }S(x))  &  =0,
\end{align*}
which after separating the real and imaginary parts and equalizing both of
them to zero, leads to the two equations%
\begin{align}
\frac{\partial}{\partial x_{\mu}}S(x)\frac{\partial}{\partial x^{\mu}}\ln
\rho(x)+\frac{1}{\rho(x)}\frac{\partial}{\partial x^{\mu}\partial x_{\mu}%
}S(x)  &  =0,\label{kgec1}\\
\frac{1}{2}\frac{\partial}{\partial x_{\mu}}S(x)\frac{\partial}{\partial
x^{\mu}}S(x)+\frac{1}{4}\frac{\partial}{\partial x^{\mu}}\ln\rho
(x)\frac{\partial}{\partial x_{\mu}}\ln\rho(x)-\frac{1}{2}\frac{1}{\rho
(x)}\frac{\partial^{2}}{\partial x_{\mu}\partial x^{\mu}}\rho(x)  &  =m^{2}.
\label{kgec2}%
\end{align}

But, equations (\ref{kgc1}) and (\ref{kgec2}) become fully equivalent to
(\ref{a}) and (\ref{b}) after assuming that the constant $\beta^{2}$ takes the
value
\begin{equation}
\beta^{2}=\frac{1}{2}.
\end{equation}

Note that this value coincides with the one before derived for the
non-relativistic situation in \cite{penacetto}. Therefore, under the defined
non-interacting approximation, it followed that the space time distribution
$\rho(x)=\int dp$ $Q(x,p)=\widetilde{Q}(x,0)$ and the also space time function
$S$ (which determines the mean value of the omentum at a given space time
point through $\langle p^{\mu}\rangle_{x}=\frac{\partial}{\partial x_{\mu}%
}S(x)$) both define a complex function $\phi(x)=(\rho(x))^{\frac{1}{2}}\exp(i$
$S(x))$ satisfying the KG equation.

In order that the searched solution of the distribution
\begin{equation}
\widetilde{Q}^{0}(x,z)=\Psi^{\ast}(z^{+})\Psi(z^{-}),
\end{equation}
can be compatible with the solution for the KG waves generating the density
$\rho(x)$ and the phase function $S$, the relation \bigskip%
\begin{align}
\widetilde{Q}^{0}(x,0)  &  =\left.  \Psi^{\ast}(z^{+})\Psi(z^{-})\right\vert
_{z\rightarrow0}\nonumber\\
&  =\Psi^{\ast}(x)\Psi(x)\nonumber\\
&  =\rho(x),
\end{align}
implies that the constant $M$ defining the ansatz, and $m$ defining the
solution of the Hamilton-Jacobi equation, should coincide.

Up to now we have discussed the generalization of the de la Pe\~{n}a-Cetto
derivation of the set of equations which should be satisfied by the joint
distribution function in the non-relativistic limit of the SQED. We also
derived a solution for the joint distribution function which implies the
satisfaction of the Klein-Gordon equation. Therefore, it rest only, in the
relativistic (as a well as in the non-relativistic one) case to check wether
or not, the derived joint distribution can obey the important positivity
property, which is required by its character as a coordinate and momenta
distribution of classical particles of mass $m,$ all showing a relativistic
momenta obeying the mass shell condition $p^{2}-m^{2}=0.$ In the coming
sections we will remark on this question.

\section{ The role of positive energy solutions}

\ Let us discuss an important physical question related with the obtained
solutions. Since the particles which are assumed to undergo the random forces,
are relativistic classical particles with rest mass $m,$ it is natural to
assume that the KG equation solution describing the stochastic motion should
be expected to be a positive energy one. This circumstance is in certain form
confirmed by a special property of the positive energy solutions (and also the
negative energy ones): their total probability conserves in time, that is
\begin{equation}
\frac{\partial}{\partial x^{0}}\int d\overrightarrow{x}\rho(x)=\frac{\partial
}{\partial x^{0}}\int d\overrightarrow{x}\phi^{\ast}(x)\phi(x)=0.
\end{equation}

This property can be easily derived after considering the following equations
also satisfied by the positive or negative energy solutions
\begin{align*}
i\frac{\partial}{\partial x^{0}}\phi(x)  &  =\pm\sqrt{m^{2}%
-(\overrightarrow{\nabla})^{2}}\phi(x)\\
&  =\pm(%
{\displaystyle\sum\limits_{m=0}^{\infty}}
(-1)^{m}\frac{((\overrightarrow{\nabla})^{2})^{m}}{2^{m}})\phi(x).
\end{align*}

For proving the condition, consider that the spatial integral of the density
for a positive energy (or a negative energy one) solution can be written and
transformed in the way%
\begin{align}
\frac{\partial}{\partial x^{0}}\int d\overrightarrow{x}\rho(x)  &
=\frac{\partial}{\partial x^{0}}\int d\overrightarrow{x}\phi^{\ast}%
(x)\phi(x)\nonumber\\
&  =\int d\overrightarrow{x}(\frac{\partial}{\partial x^{0}}(\phi^{\ast
}(x))\phi(x)-\phi^{\ast}(x)\frac{\partial}{\partial x^{0}}\phi(x))\nonumber\\
&  =\mp\frac{1}{i}\int d\overrightarrow{x}\text{ }{\Large (}(%
{\displaystyle\sum\limits_{m=0}^{\infty}}
(-1)^{m}\frac{((\overrightarrow{\nabla})^{2})^{m}}{2^{m}})\phi^{\ast}%
(x))\phi(x)-\phi^{\ast}(x)(%
{\displaystyle\sum\limits_{m=0}^{\infty}}
(-1)^{m}\frac{((\overrightarrow{\nabla})^{2})^{m}}{2^{m}})\phi(x){\Large )}%
\nonumber\\
&  =0,
\end{align}
in which there had been performed integration by parts over the all the
derivatives forming the appearing series. Thus, either the positive or
negative energy solutions define a stochastic motions conserving the total
number of particles.

This result suggests the possibility of simultaneously consider two stochastic
motions: both of them associated with positive energy solutions: but having
opposite values of the charges. Let us indicate the positive energy solutions
by $\phi_{+}(x)$ and the $\operatorname{negative}$ energy ones by\ $\phi
_{-}(x).$ But we can also define positive energy functions starting from the
negative waves by defining $\ \varphi_{+}(x)=\varphi_{+}(x^{0}%
,\overrightarrow{x})=\phi_{-}(x^{0},\overrightarrow{x}).$ In this way all the
solutions of the KG equations could participate in defining a combined
stochastic process in which two kinds of particles participate: one kind with
positive charges and the another with negative ones. We suspect that this
construction can be extended to a full covariant stochastic theory of complex
scalar particles which could appear to be a hidden variable theory for the
quantum field theory of such particles. This question will be explored in
extensions of this study. However, before considering this problem, the
approach should made consistent, by finding a positive definite joint
distribution function. The possibilities for this will be discussed in the
next section.

\section{On the positiveness of the joint distribution}

Finally, in this section we want to discuss the question about the required
positive character of the joint coordinate-momentum distribution, if at the
end the SQED approach can furnish a consistent hidden variable approach to QM.
The discussion in the present work, as the one in the original
non-relativistic analysis \cite{penacetto}, does not give a solution to this
difficulty, since again only a kind of \ Wigner distribution is derived for
the joint distribution determining the validity of the Klein-Gordon equation.
But, as it is known in general, the Wigner distributions are not positive
definite in general. Therefore, the equivalence of the SQED with QM still
remains an open question of relevant interest.

In this section below we remark about two issues related with this problem: 1)
Firstly, we search for a wavefunction $\Psi(x)$ being able to positively
define the joint distribution function in the zeroth order in the coupling
constant, by assuming that the former is defined over the mass-shell of the
random particles. The result is negative. 2) In second place, we start
exploring a possibility that SQED can be valid as describing QM effects, but
in a form compatible with the recent experimental results of Ives Couder,
showing that mechanical systems, like liquid droplets moving on oscillating
liquid surfaces, can exhibit quantum mechanical properties, as tunnel effects
and double slit interference \cite{couder,couder1}. We present a solution of
the equation for the joint distribution of SQED in the lowest order, which is
positive definite and also argue about its possible link with the recent
studies on QM mechanical properties in the movements of droplets
\cite{couder,couder1}.

\subsection{ Joint distributions defined on the mass shell.}

\bigskip Let us discuss in this subsection the point 1). The zeroth order
joint distribution had the form \ (\ref{qz})
\begin{equation}
\widetilde{Q}(x,z)=\int\frac{dp}{(2\pi)^{4}}Q(x,p)\exp(i\text{ }p^{\mu}z_{\mu
}).
\end{equation}

However, a starting hypothesis in SQED is to argue that QM can be interpreted
as a consequence of an ensemble of classical relativistic particles driven
para stochastic forces exerted on them by the vacuum. This assumption seems to
enforce that the joint distribution function must take values only within the
manifold of the phase space defined by the mass shell of the relativistic
particles $p^{2}-m^{2}=0.$ Therefore, it seems natural that the distribution
should satisfy in addition
\begin{equation}
(\frac{\partial^{2}}{\partial z^{\mu}\partial z_{\mu}}+m^{2})\widetilde{Q}%
(x,z)=0.
\end{equation}

\ But, passing from the variables $x$ and $z$ to the variables $z^{+}$ and
$\ z^{-}$, defined in (\ref{variables}), the set of two equations which should
be satisfied by the distribution can be rewritten as follows%

\begin{align}
(\frac{\partial^{2}}{\partial z^{\mu+}\partial z_{\mu}^{+}}-\frac{\partial
^{2}}{\partial z^{\mu-}\partial z_{\mu}^{-}})\text{ }\widetilde{Q}^{0}(x,z)
&  =0,\label{uno}\\
(2\beta^{2}(\frac{\partial^{2}}{\partial z^{\mu+}\partial z_{\mu}^{+}}%
-\frac{\partial^{2}}{\partial z^{\mu+}\partial z_{\mu}^{-}})+m^{2})\text{
}\widetilde{Q}^{0}(x,z)  &  =0. \label{dos}%
\end{align}

\ Let us now search for $\widetilde{Q}^{0}(x,z)\equiv$ $\widetilde{Q}%
^{0}(z^{+},z^{-})$ in the general form
\begin{equation}
\text{ }\widetilde{Q}^{0}(z^{+},z^{-})=\sum_{m,n=0}(c_{mn}\text{ }\varphi
_{m}^{\ast}(z^{+})\varphi(z^{-})+d_{mn}\varphi_{m}^{\ast}(z^{-})\varphi
(z^{+})), \label{mixed}%
\end{equation}
where the set of functions $\varphi_{m}(z^{+}),$ $m=0,1,2,...\infty$ will be
assumed to be \ orthogonal under a given positive definite scalar product.
\ Let us also impose the condition of the distribution to be real. That is%
\begin{equation}
(\widetilde{Q}^{0}(z^{+},z^{-}))^{\ast}=\widetilde{Q}^{0}(z^{+},z^{-}),
\end{equation}
which can be \ satisfied \ by choosing \
\begin{equation}
c_{mn}=d_{nm}^{\ast}.
\end{equation}

Consider the action of the operator defining the equation (\ref{uno}) on one
of the terms in the series and the requirement that each term \ satisfies
\ this equation
\begin{equation}
(\frac{\partial^{2}}{\partial z^{\mu+}\partial z_{\mu}^{+}}-\frac{\partial
^{2}}{\partial z^{\mu-}\partial z_{\mu}^{-}})\text{ }\varphi_{m}^{\ast}%
(z^{+})\varphi(z^{-})=0.
\end{equation}

Then, if the basis functions for all values of the indices $m$ satisfy \
\begin{equation}
(\frac{\partial^{2}}{\partial y^{\mu}\partial y_{\mu}}+M^{2})\text{ }%
\varphi_{m}(y)=0, \label{kg1}%
\end{equation}
each individual term in the expansion obeys the equation (\ref{uno}). \ We
will now consider the satisfaction of equation (\ref{dos}). \ After
substituting the \ expression (\ref{mixed}) \ in equation (\ref{dos}), it can
be written as follows
\begin{align}
0  &  =\sum_{m,n=0}(-2\beta^{2}M^{2}+m^{2})(c_{mn}\text{ }\varphi_{m}^{\ast
}(z^{+})\varphi(z^{-})+c_{nm}^{\ast}\varphi_{m}^{\ast}(z^{-})\varphi
(z^{+}))-\\
&  -\sum_{m,n=0}2\beta^{2}(c_{mn}\text{ }\frac{\partial}{\partial z^{\mu+}%
}\varphi_{m}^{\ast}(z^{+})\frac{\partial}{\partial z_{\mu}^{-}}\varphi
(z^{-})+c_{nm}^{\ast}\frac{\partial}{\partial z^{\mu-}}\varphi_{m}^{\ast
}(z^{-})\frac{\partial}{\partial z_{\mu}^{+}}\varphi(z^{+})).\nonumber
\end{align}

\ But, let us impose on the mass $M$ of the Klein-Gordon equation to coincide
with the mass of the particles, that is to fix $M=m$, and also substitute the
known value of $\beta^{2}=\frac{1}{2}$, which defines that (when
$\widetilde{Q}^{0}(x,0)=\Psi^{\ast}(x)\Psi(x)$) the function $\Psi(x)$ is
satisfying the KG equation. After this, the conditions for the distribution to
be defined only on the mass shell takes the form
\begin{equation}
-\sum_{m,n=0}2\beta^{2}(c_{mn}\text{ }\frac{\partial}{\partial z^{\mu+}%
}\varphi_{m}^{\ast}(z^{+})\frac{\partial}{\partial z_{\mu}^{-}}\varphi
_{n}(z^{-})+c_{nm}^{\ast}\frac{\partial}{\partial z^{\mu-}}\varphi_{m}^{\ast
}(z^{-})\frac{\partial}{\partial z_{\mu}^{+}}\varphi_{n}(z^{+}))=0.
\end{equation}

Let us assume \ now\ that $\widetilde{Q}^{0}(z^{+},z^{-})$ \ is defined by the
set of constants $c_{mn}$ given in a multiplicative form as follows \
\begin{equation}
c_{mn}=c_{m}^{\ast}c_{n\text{ }},
\end{equation}
for all values of $m$ and $n.$ This structure allows to write $\widetilde{Q}%
^{0}$ in the \ following form%
\begin{equation}
\widetilde{Q}^{0}(z^{+},z^{-})=\Psi^{\ast}(z^{+})\Psi(z^{-})+\Psi(z^{+}%
)\Psi^{\ast}(z^{-}),
\end{equation}
\bigskip and for the condition of the distribution to be defined on mass shell
it follows,%
\begin{equation}
-2\beta^{2}(\frac{\partial}{\partial z^{\mu+}}\Psi^{\ast}(z^{+})\frac
{\partial}{\partial z_{\mu}^{-}}\Psi(z^{-})+\frac{\partial}{\partial z^{\mu-}%
}\Psi^{\ast}(z^{-})\frac{\partial}{\partial z_{\mu}^{+}}\Psi(z^{+}))=0.
\end{equation}
The above relation should be satisfied for all values of $z^{+}$ and $z^{-}$.
Thus making $z^{+}=z^{-}=x$ \ and expressing the solution of a positive energy
KG equation as
\begin{align}
\Psi(x)  &  =\int\frac{d\overrightarrow{p}}{(2\pi)^{3}p^{0}}%
c(\overrightarrow{p})\exp(-i\text{ }p\, x),\\
p^{0}  &  =\sqrt{m^{2}+\overrightarrow{p}^{2}},\nonumber
\end{align}
and integrating over the three spatial coordinates \ it follows
\begin{align}
0  &  =-4\beta^{2}\int d\overrightarrow{x}\frac{\partial}{\partial x^{\mu}%
}\Psi^{\ast}(x)\frac{\partial}{\partial x_{\mu}}\Psi(x)\nonumber\\
&  =-4\beta^{2}\int\frac{d\overrightarrow{p}}{(2\pi)^{3}p^{0}}\frac
{d\overrightarrow{q}}{(2\pi)^{3}q^{0}}c(\overrightarrow{p})c^{\ast
}(\overrightarrow{q})(p^{0}q^{0}-\overrightarrow{p}.\overrightarrow{q}%
)\times\nonumber\\
&  \int d\overrightarrow{x}\exp{\LARGE (}-i{\Large (}(p^{0}-q^{0}%
)x^{0}-(\overrightarrow{p}-\overrightarrow{q}).\overrightarrow{x}%
{\Large )}{\LARGE )}\nonumber\\
&  =-4\beta^{2}\int\frac{d\overrightarrow{p}}{(2\pi)^{3}p^{0}}\frac
{d\overrightarrow{q}}{q^{0}}c(\overrightarrow{p})c^{\ast}(\overrightarrow{q}%
)(p^{0}q^{0}-\overrightarrow{p}.\overrightarrow{q})\times\nonumber\\
&  \delta^{(3)}(\overrightarrow{p}-\overrightarrow{q})\exp{\large (}%
-i(p^{0}-q^{0})x^{0}{\large )}\nonumber\\
&  =-4\beta^{2}\int\frac{d\overrightarrow{p}}{(2\pi)^{3}(p^{0})^{2}%
}c(\overrightarrow{p})c^{\ast}(\overrightarrow{q})(m^{2}+\overrightarrow{p}%
^{2}-\overrightarrow{p}.\overrightarrow{p})\nonumber\\
&  =-4\beta^{2}m^{2}\int\frac{d\overrightarrow{p}}{(2\pi)^{3}(p^{0})^{2}%
}c(\overrightarrow{p})c^{\ast}(\overrightarrow{q}).
\end{align}

The obtained relation implies that the $c(\overrightarrow{p})$ function should
vanish, and thus it follows that no function $\Psi(x)$ satisfying the
Klein-Gordon equation (and the few restrictive conditions being assumed, like
: $m=M$ and $c_{mn}=c_{m}^{\ast}c_{n\text{ }}$\ ) can be used to construct a
positive definite joint coordinate and momentum distribution function defined
as
\begin{equation}
\widetilde{Q}^{0}(z^{+},z^{-})=\Psi^{\ast}(z^{+})\Psi(z^{-})+\Psi(z^{+}%
)\Psi^{\ast}(z^{-}),
\end{equation}
for the description of a stochastic motion of relativistic classical particles
which should have their 4-momentum well defined on the mass shell
$\ p^{2}-m^{2}.$

\subsection{A positive joint distribution based in localized particles: a link
with Couder's experiments?}

In this section we will present a solution of the equation for the joint
distribution $Q(x,p)$ being positive definite in phase space. It is based in
the Yukawa potential like solution of the Klein-Gordon equation in the
presence of sources
\begin{align}
\psi(\overrightarrow{x})  &  =\frac{1}{4\pi}\frac{\exp(-m\text{ }r)}{r},\\
r  &  =\sqrt{x_{1}^{2}+x_{2}^{2}+x_{3}^{2}},
\end{align}
satisfying the KG equation
\begin{equation}
(\partial^{2}-m^{2})\psi(\overrightarrow{x})=-\delta^{(3)}(\overrightarrow{x}%
),
\end{equation}
where $\delta^{(3)}$ is the three dimensional Dirac delta function. After
performing a Lorentz transformation to a frame moving with velocity
$\overrightarrow{v}$ along, let say, the $x_{1}$ axis$,$ followed by a shift
in the position of the origin of coordinates (at vanishing time $t$) to an
arbitrary point $\overrightarrow{x}_{k},$ this Yukawa like solution becomes a
"moving" one, of the form \
\begin{align}
\psi_{x_{k},}(x,p)  &  =\frac{1}{4\pi}\frac{\exp(-m\text{ }\sqrt
{\frac{\overrightarrow{p}^{2}+m^{2}}{m^{2}}(x_{1}-\frac{p_{1}}{\sqrt
{\overrightarrow{p}^{2}+m^{2}}}t-x_{1}^{k})^{2}+(x_{2}-x_{2}^{k})_{2}%
^{2}+(x_{3}-x_{3}^{k})})}{\sqrt{\frac{\overrightarrow{p}^{2}+m^{2}}{m^{2}%
}(x_{1}-\frac{p_{1}}{\sqrt{\overrightarrow{p}^{2}+m^{2}}}t-x_{1}^{k}%
)^{2}+(x_{2}-x_{2}^{k})_{2}^{2}+(x_{3}-x_{3}^{k})}}\label{lump}\\
x  &  =(t,\overrightarrow{x}),\text{ \ \ \ }p=(p^{0},\overrightarrow{p}%
)=(\frac{m}{\sqrt{1-v^{2}}},\frac{m\text{ }\overrightarrow{v}}{\sqrt{1-v^{2}}%
}),\text{ \ \ }x_{k}=(0,\overrightarrow{x}_{k}).
\end{align}

In this change it had been used the relations
\begin{equation}
v=\frac{p_{1}}{\sqrt{\overrightarrow{p}^{2}+m^{2}}},\text{ \ }\sqrt{1-v^{2}%
}=\frac{m}{\sqrt{\overrightarrow{p}^{2}+m^{2}}}.
\end{equation}
This function also can be expressed in a rotational invariant form as%

\begin{equation}
\psi_{x_{k},}(x,p)=\frac{1}{4\pi}\frac{\exp(-m\text{ }\sqrt{\frac
{\overrightarrow{p}^{2}+m^{2}}{m^{2}}{\Large (}(\overrightarrow{x}%
-\frac{\overrightarrow{p}}{\sqrt{\overrightarrow{p}^{2}+m^{2}}}%
t-\overrightarrow{x}^{k}).\frac{\overrightarrow{p}}{|\overrightarrow{p}%
|}{\Large )}^{2}+(\overrightarrow{x}-\overrightarrow{x}^{k}%
).(\overrightarrow{\overrightarrow{I}}-\frac{\overrightarrow{p}%
\overrightarrow{p}}{|\overrightarrow{p}|^{2}}).(\overrightarrow{x}%
-\overrightarrow{x}^{k})})}{\sqrt{\frac{\overrightarrow{p}^{2}+m^{2}}{m^{2}%
}{\Large (}(\overrightarrow{x}-\frac{\overrightarrow{p}}{\sqrt
{\overrightarrow{p}^{2}+m^{2}}}t-\overrightarrow{x}^{k}).\frac
{\overrightarrow{p}}{|\overrightarrow{p}|}{\Large )}^{2}+(\overrightarrow{x}%
-\overrightarrow{x}^{k}).(\overrightarrow{\overrightarrow{I}}-\frac
{\overrightarrow{p}\overrightarrow{p}}{|\overrightarrow{p}|^{2}}%
).(\overrightarrow{x}-\overrightarrow{x}^{k})}} , \label{lump1}%
\end{equation}
where $(\overrightarrow{\overrightarrow{I}}-\frac{\overrightarrow{p}\text{
}\overrightarrow{p}}{|\overrightarrow{p}|^{2}})$ is the projection tensor on
the plane orthogonal to the velocity and $\frac{\overrightarrow{p}%
}{|\overrightarrow{p}|}$ is a unit vector in the direction of the velocity.

The function $\psi_{x_{k},}(x,p)$ is positive definite in the phase space
$(x,p).$ Also, in the rest frame $p=(m,\overrightarrow{0}),$ it directly
satisfies the equation for the joint momenta-coordinate distributions, since
it is time independent and the three velocity vanishes. Then%
\begin{equation}
p^{\mu}\frac{\partial}{\partial x^{\mu}}\psi_{x_{k},}(x,p)=0.
\end{equation}

But, having this equation a covariant form, it should be also valid after
performing any Lorentz transformation of the coordinates and momenta. Now, we
can define a set of $N$ points $x_{k},k=1,2,....,N$ . Then, by superposing the
functions of the type (\ref{lump1}) for all the values of $k$, more general
solutions can be constructed. They will describe a set of $N$ localized
solutions of the Klein-Gordon equation with sources. Also, the new
distributions also will be positive quantities in all the mass-shell phase
space. These solutions of the equations $p^{\mu}\frac{\partial}{\partial
x^{\mu}}Q(x,p)=0,$ showing a positive joint distribution have the forms
\begin{equation}
Q(x,p)=\sum_{k}\psi_{x_{k},}(x,p),
\end{equation}
where for each $k$ value, the velocity for the distribution with index $k$,
points in an arbitrary direction.

Let us now argue about the possible links of these special joint distributions
with the Couder experimental results \cite{couder,couder1}. For this purpose,
let us qualitatively discuss a situation in which one of the solution
$\psi_{x_{k},}(x,p)$ with a given velocity, perpendicularly approaches a wall
having two slits holes. Let us first note that the singularity of the solution
is similar to the one in the Coulomb potential field and its classical energy
is infinite. Therefore, since we will assume that the particle has a finite
mass $m,$ some negative contribution to the bound energy (which is
concentrated in the singularity point) should cancel the infinite and positive
contribution of the classical energy outside the point. Then, let us consider
a sphere centered in the singularity at any instant, such that the field
energy outside the sphere coincides with the total mass of the particle.
Therefore, the contribution to the total energy inside the defined sphere,
should vanish (the negative cohesive term should cancel the infinite positive
energy laying outside the point, but inside the sphere). However, having not
net mass, the system inside the sphere, might be suspected to weakly
contribute to the free dynamics of the particle. Then, we have that outside
the small sphere the system will satisfy the Klein-Gordon equation and also
will move as a whole with constant velocity, Thus, the solution $\psi_{x_{k}%
,}(x,p)$ might perhaps also be approximately represented by a wave packet
solving KG equation, but with a momenta distribution showing non vanishing
values only in a small neighborhood of the momenta component $p_{1}=\frac
{mv}{\sqrt{1-v^{2}}}$, in order that the packet shows a constant velocity. If
this idea is valid, then
\begin{align}
\psi_{x_{k},}(x,p) &  \simeq\int d\overrightarrow{p}\text{ }%
f(\overrightarrow{p})\exp(-i\sqrt{\overrightarrow{p}+m^{2}}%
t+\overrightarrow{p}.\overrightarrow{x}\text{ }i)\nonumber\\
&  \simeq\int dp_{1}dp_{2}dp_{3}\text{ }f(\overrightarrow{p})\exp(-i\text{
}p(\frac{\sqrt{\overrightarrow{p}+m^{2}}}{p}t-x_{1})+i\text{ }p_{2}%
x_{2}+i\text{ }p_{3}x_{3})\nonumber\\
&  \simeq\int dp_{1}dp_{2}dp_{3}\text{ }f(\overrightarrow{p})\exp(-i\text{
}p(v\text{ }t-x_{1})+i\text{ }p_{2}x_{2}+i\text{ }p_{3}x_{3}).\label{wp}%
\end{align}

However, it should be noted that the singular solution is localized  within
region of the size  of a Compton associated to the free mass parameter $m$.
This is a small quantity for usual particles as, by example electrons. Thus,
in order that the representation  ($\ref{wp}$) could be valid,  the momentum
bandwidth of the integral in ($\ref{wp}$) should be larger than one over the
spatial width of the singular solution $\delta x\simeq\frac{1}{m},$ that is,
larger than $m.$  In the relativistic limit $p_{1}>>m$, this condition can be
satisfied. However, for mass parameters $m$ larger than the electron's
one, and in the non relativistic limit, this rule can not be
imposed. Thus, we  expect  that the parameter $m$ should have a small
value in order to allow implementing  the representation ($\ref{wp}$) in the
non relativistic limit. Since the Couder's experiments are done for
movements of the droplets over surfaces having massless propagating modes,
even the vanishing mass parameter $m$  could be allowed.

To end the argue, note that ($\ref{wp}$) is a wave corresponding to particles
of mass $m$ and wavelength $\lambda=\frac{2\pi}{p}.$ Therefore, assumed that
the two slits to which the wave approaches, have a separation of a similar
size to the wavelength $\lambda$, these modes should tend to be scattered by
the action of the slits. Clearly, if the waves were completely free ones, this
scattering should work. However, since the singularity is expected to maintain
the structure of the particle when the scattering process occurs (as it
happens in the Couder's experiments) the whole effect of such "dispersion
forces" could be suspected to be reduced to control the movement of the
singularity (if it passes through the slits) to be pointing in the directions
of the usual interference maxima. This argue indicates that the results of
Couder could perhaps be described by the found positive solutions of the
equation for the joint distribution function. The study of the two slits
scattering on the here defined singular solutions will be considered elsewhere.

\section*{SUMMARY}

We have presented a generalization of the non-relativistic stochastic quantum
mechanics introduced by de la Pe\~{n}a and Cetto \cite{penacetto}. The
discussion starts form the description of the random motions of a particle
under the action of a relativistically invariant stochastic force defined in
reference \cite{boyer} \ It is checked that in the non relativistic limit the
starting equations reduce to the ones employed in \cite{penacetto}. Then, the
set of equations for the joint distribution is expanded as a series of the
particle charge. The free approximation, that is, the equation following in
the zero order of the expansion in the charge is solved by considering natural
ansatz for the distribution. After this, it is argued that the space-time
probability distribution of the stochastic process $\rho(x)$ and a phase
function $S(x)$ which gradient determines the momentum mean value at a
definite space-time point, define a complex scalar function satisfying the KG
equation through $\phi(x)=\sqrt{\rho(x)}\exp(i$ $S(x)).$ It is argued that the
total number of particles $N=\int d\overrightarrow{x}$ $\rho(x)$ determined by
the space-time distribution conserves in time, if the KG solutions considered
for determining the distributions are either positive or negative energy
modes. One important consistency question of SQED is also discussed: the
possibility that the joint coordinate and momenta distribution function can
really be definite positive in the mass shell manifold of the relativistic
classical particles \ $p^{2}-m^{2}=0.$ The particular analysis presented gave
a negative result. That the particular solution found for the zeroth order
equation for the joint distribution function, can not define a proper on mass
shell distribution function cast some doubts about the possibility of being
able to show the equivalence of the quantum mechanics or quantum field theory
with SQED. However, there is a special circumstance that seem could help to
overcome this limitation: The KG equation, when seen as theory of particles
shows the interesting effect that it lacks a standard definition for the
position operator eigenfunction as the Dirac's Delta function. In place of it,
the appropriate position operator for this theory has a Gaussian like
appearance showing non vanishing values in spatial regions of the size of the
Compton wavelength of the particles\cite{schweber}. Therefore, it looks
reasonable that the proper interpretation of the $\widetilde{Q}^{0}(x,p)$\ as
describing particles with a well defined position $x$ can show difficulties,
as they followed in the here presented discussion. This leads to the idea that
a proper modification of the kinetic equations to take account of this
extensive nature of the particles occassionally could still lead to a
consistent equivalence of the SQED with QM and QFT.

The work is also presenting solutions for the joint distribution functions
which obey the positive conditions required by a proper classical
distribution. It is also argued that these solutions could be related with
systems of particles showing the Couder's experimental results, if the
localized solutions employed for their construction can be approximately
represented as massive wavepackets \cite{couder}. The investigation of this
possibility is expected to be considered elsewhere.

Assumed that the identified difficulties with positive condition of the joint
distribution can be properly solved, the discussion opens possibilities for
the extension of the SQED analysis. Of particular interest are: 1) The
possibility of extending the discussion to describe the stochastic evolution
of two sets of particles: one described by the positive energy solutions and
the other by the negative energy ones. After to also including the presence of
an external electromagnetic fields, this construction seem to offer
opportunities for describing the creation and annihilation particles by the
electromagnetic field. The interaction terms, could result to be sources of
the variation in the total numbers of positive or negative charged particles
generated by the annihilation or creation of particles due to the action of
the electromagnetic field. The search for the connection of this construction
with the quantum field theory of the complex scalar field is a further
question of interest to explore. 2) Another issue which is suggested by the
discussion is the possibility, that in the same way that a two sets of
positive and negative charged scalar particles could be described, it could be
possible to search for set of spinor waves which densities and phase functions
could also define stochastic motions of classical particles showing spin
$\frac{1}{2}$.

\section*{ACKNOWLEDGMENTS}

The support granted by the N-35 OEA Network of the ICTP is greatly appreciated.

\end{document}